\documentclass[smallextended]{svjour3}       % onecolumn (second format)
\smartqed  % flush right qed marks, e.g. at end of proof
\usepackage{graphicx}
\usepackage{amssymb,amsmath}
\def\makeheadbox{\relax}

\usepackage[hidelinks]{hyperref}
\hypersetup{
	unicode=false,          % non-Latin characters in Acrobat’s bookmarks
	pdftoolbar=false,        % show Acrobat’s toolbar?
	pdfmenubar=false,        % show Acrobat’s menu?
	pdffitwindow=false,     % window fit to page when opened
	pdfstartview={FitH},    % fits the width of the page to the window
	pdftitle={ Optimal investment to enable evolutionary rescue },    % title
	pdfauthor={  Ashander  Thompson  Sanchirico  Baskett  },     % author
	colorlinks=true,       % false: boxed links; true: colored links
	linkcolor=red,          % color of internal links
	citecolor=blue,        % color of links to bibliography
	filecolor=blue,      % color of file links
	urlcolor=blue           % color of external links
}

\title{Optimal Investment to Enable Evolutionary Rescue}

\author{Jaime Ashander \and
	Lisa C. Thompson \and
	James N. Sanchirico \and
	Marissa L. Baskett
}
\authorrunning{Ashander, Thompson, Sanchirico, and  Baskett}
\institute{Jaime Ashander \at
	Center for Population Biology, University of California---Davis, Davis, CA 95616\\
	Department of Environmental Sciences and Policy, University of California---Davis, Davis, CA 95616\\
        \href{https://orcid.org/0000-0002-1841-4768}{ORCID: 0000-0002-1841-4768} \\
	\email{jashander@ucdavis.edu}            \\
	\emph{Present address:} Rsources for the Future, 1616 P St NW, Washington DC 20036
\and
	Lisa C. Thompson \at
	Department of Wildlife, Fish, and Conservation Biology, University of California---Davis, Davis, CA 95616\\
	Regional San (Sacramento Regional County Sanitation District) and Sacramento Area Sewer District 10060 Goethe Road, Sacramento, CA 95827
\and
	James N. Sanchirico \at
	Department of Environmental Sciences and Policy, University of California---Davis, Davis, CA 95616
\and
	Marissa L. Baskett \at
	Center for Population Biology, University of California---Davis, Davis, CA 95616\\
	Department of Environmental Sciences and Policy, University of California---Davis, Davis, CA 95616
}

\begin{document}
\maketitle

\begin{abstract}
``Evolutionary rescue'' is the potential for evolution to enable population % MLB edit: added "to" before enable
persistence in a changing environment. Even with eventual rescue,
evolutionarily time lags can cause the population size to temporarily
fall below a threshold susceptible to extinction. To reduce extinction
risk given human-driven global change, conservation management can
enhance populations through actions such as captive breeding. To
quantify the optimal timing of, and indicators for engaging in,
investment in temporary enhancement to enable evolutionary rescue, we
construct a model of coupled demographic-genetic dynamics given a moving
optimum. We assume ``decelerating change'', as might be relevant to
climate change, where the rate of environmental change initially exceeds
a rate where evolutionary rescue is possible, but eventually slows. We
analyze the optimal control path of an intervention to avoid the
population size falling below a threshold susceptible to extinction,
minimizing costs. We find that the optimal path of intervention
initially increases as the population declines, then declines and ceases
when the population growth rate becomes positive,
%% MLB edit (addition):
which lags the stabilization in environmental change.
In other words, the
optimal strategy involves increasing investment even in the face of a
declining population, and positive population growth could serve as a
signal to end the intervention. In addition, a greater carrying capacity
relative to the initial population size decreases the optimal
intervention. Therefore, a one-time action to increase carrying
capacity, such as habitat restoration, can reduce the amount and
duration of longer-term investment in population enhancement, even if
the population is initially lower than and declining away from the new
carrying capacity.
\end{abstract}

\section{Introduction}\label{introduction}

Global environmental change such as climate change has the potential to
exceed the physiological tolerances of many organisms (Thomas et al.
2004; Urban 2015). For a population faced with environmental conditions
outside its range of tolerance, persistence might occur through either a
shift in its range or genetic adaption (Davis et al. 2005). Persistence
via genetic adaptation in response to environmental change in a
population that would otherwise perish is called ``evolutionary rescue''
(ER, Gomulkiewicz and Holt 1995; Carlson et al. 2014).

To date, theory on evolutionary rescue has focused on two situations
where it can occur naturally. First, if the environmental optimum shifts % MLB edit: added comma after Frist
suddenly, population size initially declines and eventually increases if
enough genetic variation relative to the amount of change exists for
adaptation to the new environment to occur (Gomulkiewicz and Holt 1995;
Carlson et al. 2014). Such evolutionary rescue typically involves a
period of low population size during which a population might be
susceptible to factors such as demographic stochasticity, environmental
stochasticity, Allee effects, inbreeding, and genetic drift % that can
% drive extinction in small populations
(Lande 1998; Gilpin and Soule 1986).  % MLB edit (x2): added year on Gilpin and Soule citation, commented out redundant phrase at the end of the sentence
Second, if the environmental optimum is continuously changing at a
constant rate, population growth declines, but populations with enough
genetic variance relative to the rate of environmental change maintain
population growth (Lynch and Lande 1993; B\"{u}rger and Lynch 1995).
Therefore, populations with a given amount of genetic variation have a
``critical rate'' of environmental change above which ER cannot occur
and growth rates become negative (Kopp and Matuszewski 2013).

As an example of a changing environmental optimum, climate change lies
between sudden shift and gradual change. Depending on the amount of
greenhouse gas emissions and therefore the rate of change in the climate (e.g.,
mean annual temperature), there % MLB edit: added "the" before rate of climate change
might be a period of time where the rate of change in the optimum is
``super-critical'', exceeding the rate where evolutionary rescue can
occur. % Eventually, h % MLB edit: deleted 1st eventually so not twice in the same sentence
However, as the rate of change decelerates, as
eventually occurs for all future climate scenarios (Meinshausen et al.
2011), evolution might play a greater role in population persistence.

Conservation management to increase the likelihood of evolutionary
rescue and therefore population persistence under environmental change
such as climate change can take two forms: mitigation and adaptation.
Mitigation to reduce the rate or amount of change in temperature (e.g., by
reducing greenhouse gas emissions) can increase the ability
for evolution to keep up with the changing environment. However, for
climate change, mitigation requires international
cooperation (King 2004). Conservation management, however, most often
occurs at local, regional, or national scales.
Further, local efforts to mitigate emissions do not reduce locally-felt effects of climate change.
Without a direct role in mitigating climate change, then, conservation management must focus on ``adaptation''
in the anthropogenic sense, which in a conservation context involves
promoting processes that increase the likelihood of population
persistence (Stein et al. 2013). For the case of increasing the
likelihood of ER, adaptation can involve reducing local stressors (e.g.,
Baskett et al. 2010) or enhancing population size to reduce the
likelihood of a population falling below a threshold size at risk of
extinction (Fraser 2008).

For decelerated change such as climate change, management interventions
during the initial period when change might be super-critical could
preserve the option for longer-term ER to occur. Interventions inevitably differ in
whether they have a temporary or permanent effect on population size and
growth rate. Interventions with potentially permanent effects
include habitat restoration (Bradshaw 1996) and removal of invasive
predators or competitors (Myers et al. 2000).  Interventions with temporary
effects, i.e., which only affect the population transiently, include resource provisioning
(Ruffino et al. 2014), head-starting (captive rearing of a vulnerable
early life stage), and captive breeding (Heppell et al. 1996; Griffiths
and Pavajeau 2008). Climate change threatens a variety of species that are also
targets for captive breeding. For example, climate change-driven changes to river flow and temperature can negatively affect
Pacific salmon (Crozier et al 2008), and hatcheries (i.e., hatching of eggs in captivity to release into the wild at early life stages)
are a long-standing tool to increase salmon population sizes (Naish et al 2008).
Analogously, increases in extreme temperature events threaten the persistence of tropical corals (Bellwood et al 2004),
and "coral gardening" (i.e., nursery-based growth of small fragments into larger corals to outplant into the wild)
can provide large-scale population supplementation for corals (Lirman and Schopmeyer 2016).
Yet, captive rearing and breeding have the potential to
involve unintended negative consequences for wild populations such as
domestication, the negative effects of which can accumulate over
multiple generations, which leads to recommendations to limit the use and
duration of such programs (Snyder et al. 1996; Fraser 2008). In addition,
the ultimate success of captive breeding and rearing in leading to
population persistence without requiring indefinite intervention (i.e.,
conservation reliance \emph{sensu} Scott et al. 2010), depends on
addressing the factors that originally lead to population declines
(Fraser 2008).

In addition to the potential to incur unintended consequences,
interventions such as captive breeding and rearing can be costly (Snyder
et al. 1996) and budgets are inevitably limited. Thus, a key management
question is the efficient allocation of resources both over time and among
populations. For example, when is it bioeconomically optimal to invest in an
intervention and for how long should a manager keep investing?
Furthermore, what biological or economic indicators can be used to make such
decisions? Investing early may help build population
abundance and reduce the effects of environmental change. Alternatively, for % MLB edit: added "the" before effects
populations initially close to carrying capacity and thus
self-regulating, early investments may have less effect per dollar
spent. Self-regulation might also determine the efficacy of pairing an
investment with a temporary effect such as captive rearing with an
action with a permanent effect such as habitat restoration. In
particular, a one-time investment to permanently increase carrying
capacity might reduce the investment necessary in captive rearing by
decreasing the role of self-regulation, or it might have little effect
if self-regulation has little effect on population dynamics when
populations are initially declining under rapid environmental change.
Economic factors that might further influence the pattern of investment
include budget constraints and the rate of discounting. Possible
indicators for optimal timing of investment include a population growth
rate, population size, or the rate of environmental change.

\begin{figure}[p!]
  \centering
  \includegraphics[width=\textwidth]{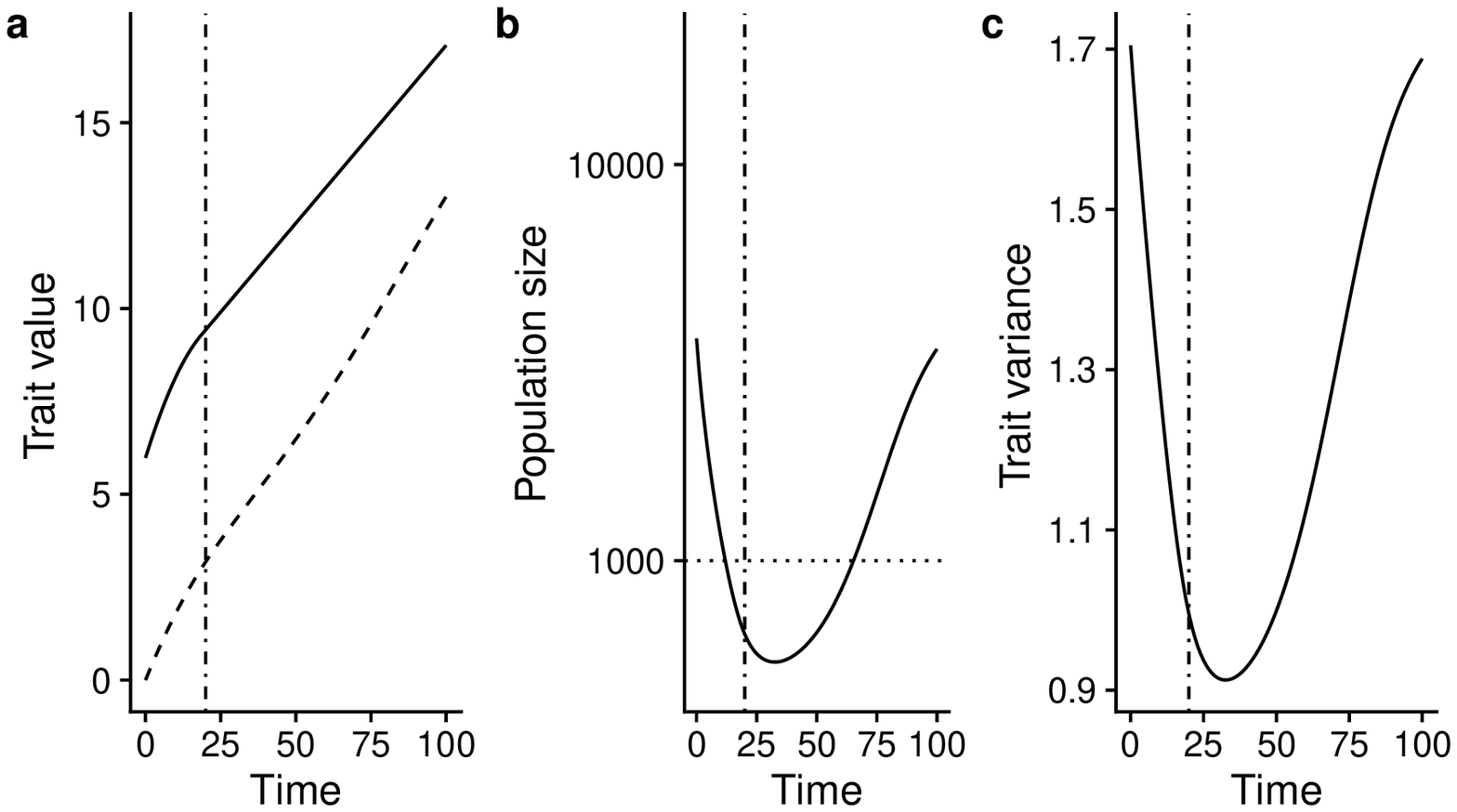}
	\caption[Change in environmental optimum, trait value, and population size.]{Under ``decelerating environmental change'' (a), the optimum trait value
(solid line) initially increases rapidly, then slows to the
critical rate where evolutionary rescue can occur at time
$t_\text{safe}$ (vertical dash-dot line). The population mean trait
(dashed line) initially lags % further and further % MLB edit: reduce extraneous words
from the optimum but
after $t_\text{safe}$ closes the gap.
Without intervention to
supplement or improve population growth (b), the population will fall
below critically low size susceptible to extinction (gray line) for an
extended period, but it does eventually increase.
Meanwhile, the genetic variance decreases with decreasing
population size (c) according to the stochastic house-of-cards approximation.
}
	\label{figureenv}
\end{figure}

Here we quantify the bioeconomically optimal investment schedule for an
evolving population undergoing decelerating environmental change.
The objective of the regulator is to minimize costs (and therefore the amount of intervention) given a goal
of avoiding extinction. To this end, we develop a model that couples the
demographic dynamics necessary to account for extinction risk, the
genetic dynamics necessary to account for ER, and the economic dynamics
necessary to determine the optimal investment schedule. Our biological
model assumes a moving optimum where the rate initially exceeds the
critical rate for ER to occur and eventually slows to that rate (Figure
\ref{figureenv}a). Without intervention the population size will decline
below a critical threshold considered at risk of extinction (Figure
\ref{figureenv}b). We also assume a management intervention that
temporarily increases population growth (e.g.~resource provisioning,
head-starting, or captive breeding) but is costly. We analyze the
pattern of intervention that minimizes costs, subject to the constraint
of keeping the population above a critical size, given different values
for the carrying capacity, discount rate, and annual budget.
% MLB edit: deleted mention of different time scales of budget constraints given that we now only present the results with annual constraints; added "annual" before budget
%, either annually or
% in total.

\section{Materials and Methods}\label{materials-and-methods}

Our bioeconomic model consists of a sub-model for the environment, the
biological response of the population, and the economic costs of control
(i.e., management interventions to improve population growth). Combining
these sub-models, we pose an control problem for optimally-scheduling
spending on the control while avoiding extinction. We analyze the
problem numerically to find the optimal solution. % MLB comment: is "non-linear program" a commonly-accepted term for this approach?  If not, frame in terms of doing an optimal control analysis?

\subsection{Model}\label{model}

\subsubsection{Changing environment}\label{changing-environment}

To represent environmental change, we consider an environmental optimum
$\theta(t)$ that changes in time with rate $k(t)$. Initially, the
rate of change $k_0$ exceeds a critical rate $k_c$, above which
evolution cannot prevent population declines (e.g., as in Lynch and
Lande 1993) but it slows to less than % around % MLB edit: replaced "around" with "less than"
$k_c$ by time $t_\text{safe}$. We
assume the optimum changes deterministically,
\begin{equation}\label{E:env}
\theta (t)  = k(t) t % + \varepsilon
% MLB edit: commented out \varepsilon because we're no deterministic
% MLB comment: is this correct?
\end{equation}
where
\begin{equation}\label{E:kappa} k(t) = \begin{cases} k_c \left(\kappa_0 - \frac{\kappa_0 -
\kappa_\text{min}}{t_\text{safe} - t_0} t \right) \quad \text{for}  \: t < t_\text{safe} \\ \kappa_\text{min} k_c \quad
\text{for}   \: t \ge t_\text{safe} \end{cases}, \end{equation} % MLB edit: moved the comma to be after the equation rather than in the text that follows
and with $\kappa_0 > 1$ and $\kappa_\text{min} < 1$.

\subsubsection{Biological dynamics}\label{biological-dynamics}

Our model follows the joint demographic-genetic dynamics of population
size $N(t)$ and genetic distribution $\psi_t(a)$ of quantitative
trait $a$ under stabilizing selection toward the optimum
$\theta(t)$. We assume the order of events in the life-cycle is
mating, density-dependence, then viability selection. Note our
life-cycle ordering corresponds to hard selection (Wallace 1975). We
also assume random mating, a closed population, and discrete
generations. Finally, we assume many genes of small effect additively
contribute to the genotype such that, by the central limit theorem, the genetic distribution $\psi(a)$ is normal (Lande 1976).
Therefore, we define the genetic distribution by its
evolving mean $\bar{a}_t$ and genetic variance $\sigma^2_a(N)$,
which depends on the census population size to accuont for the effects of drift, % depending on the census population size % MLB edit: moved this below
$\psi_t(a)=\exp(-(a-\bar{a}_t)^2/(2\sigma^2_a(N)))/\sqrt{2\pi\sigma^2_a(N)}$. % MLB edit: added (N) to \sigma^2_a's
Specifically we use
the stochastic house of cards approximation
% MLB edit: addition for explanation of what this approximation is
of mutation-selection-drift balance for the genetic variance $\sigma_a$,
as in B\"{u}rger and Lynch
(1995), which we specify below.  % and which partially accounts for the % MLB edit: moved this above
% effects of declining population size on genetic variance.

In the mating step, the number of offspring per individual is $R_0$,
and the assumption of random mating % and constant genetic variance  % MLB edit: deleted phrase because genetic variance isn't constant with house-of-cards approximation (note also assumption then singular instead of plural)
means
that the genetic distribution is unchanged (Lande and Arnold 1983). In
the density dependence step, we apply a saturating Beverton-Holt (1957)
function with parameter $K$ determining carrying capacity (equal to
$(R_0 - 1)K$), where density-dependent survival is independent of
genotype. Therefore, encapsulating both reproduction and density
dependence, the
% MLB edit: addition to recognize that selection also modifies growth
pre-selection
growth function
$g(N(t)) = \frac{R_0 N(t)}{1 + N(t)/K}$ depends solely on the
population size $N(t)$.

In the viability selection step, we convert genotype $a$ to phenotype
$z$ given random environmental contribution to the phenotype $e$
normally distributed with mean 0 and variance $\sigma_e^2$, i.e.~we
account for imperfect inheritance but not phenotypic plasticity, such
that $z = a + e$. Therefore, the phenotype probability distribution
given a particular genotype is
$P(z|a)=\exp(-(z-a)^2/(2\sigma_e^2))/\sqrt{2\pi\sigma_e^2}$. We then
apply stabilizing selection for $\theta(t)$ given width of the fitness
function (inverse of selection strength) $\omega^2$, such that % MLB edit: removed 1/ from \omega^2 here, because \omega^2 is the width of the fitness function in the latest formulation (I think it was inverted before)
fitness $W(z)=\exp(-(z-\theta(t))^2/\omega^2$. Applying selection to
the genetic distribution yields the genotypic distribution at time $t$ as % MLB edit: changed phenotype to genotype
$\psi'_t(a) = \int W(z)P(z|a)\psi_t(a)dz$, where the overall % MLB comment: if the \psi equation still runs over the page when complied, start a new line here
population fitness in generation $t$, equivalent to the proportion of the population that
survives viability selection, is
\begin{equation}\label{E:Wbar}
\bar{W}(t)=\int\psi'_t(a)da=\sqrt{\frac{\omega^2}{\omega^2+\sigma_a^2(N)+\sigma_e^2}}e^{-\frac{(\bar{a}_t-\theta(t))^2}{2(\omega^2+\sigma_a^2(N)+\sigma_e^2)}}.
\end{equation}
Therefore, as $\theta(t)$ changes each generation, mean fitness changes as well, cascading into changes in the population size and genetic distribution.
For the population size, applying fitness-dependent survival after growth yields the recursion of $N(t+1)=\bar{W}(t)g(N(t))$.
Using the above-described growth function that accounts for reproduction and density dependence, the overall natural population growth factor (excluding any intervention-based growth), calculated from $N(t+1)/N(t)$, is:
\begin{equation}\label{E:lambda}
\bar{\lambda}(t,N) =  \frac{\bar{W}(t)R_0}{1+N(t)/K}. % MLB comment: please double-check this to make sure my formulation is correct
\end{equation}

For the genetic dynamics, we normalize the genetic distribution
$\psi_{t+1}(a)=\psi'_t(a)/\bar{W}(t)$ to yield the new genotypic % MLB edit: changed phenotypic to genotypic
distribution with mean
\begin{equation}\label{E:arecursion}
\bar{a}_{t+1} = \frac{\theta(t)\sigma_a^2(N)+(\omega^2+\sigma_e^2)\bar{a}_t}{\omega^2+\sigma_a^2(N)+\sigma_e^2}. % MLB edit: removed ( before \theta(t)
\end{equation}
To simplify notation, we let
$s(N)=\sigma_a^2(N)/(\omega^2+\sigma_a^2(N)+\sigma_e^2)$ and rearrange
to arrive at the mean genotype recursion
\begin{equation}\label{E:arecursion2}
\bar{a}_{t+1} = \bar{a}_{t}+s(\theta(t)-\bar{a}(t)).
\end{equation}

In these recursions we use the stochastic house of cards (SHC) approximation
as in B\"{u}rger and Lynch (1995):
first, setting the effective population size to
$N_e(N) = \frac{2 R_0 }{2 R_0 - 1} N$ and, second, using the formula % MLB edit (x2): 0 subscripted for R_0
$\sigma_a^2(N) = \frac{2 V_m N_e(N) }{ 1 + \alpha^2 N_e(N) /(\omega^2 + \sigma_e^2)}$, %.
% MLB edit: addition to define all parameters used; definitions take from Table 1
where $\alpha^2$ is the genetic effect size variance of a new mutation and $V_m$ is the mutational variance.
The SHC approximation accounts for the equilibirium effect of changing population size on genetic
variance with a fixed optimum, constant mutational variance, effect size, and demography;
using it for dynamic population size change as we do (consistent with B\"{u}rger and Lynch 1995)
is inexact but caputres the coarse-scale effect of population size change on genetic variance (Kopp and Matuszewski 2013).

Our model for a decelerating optimum, eq.~\eqref{E:env}, requires choosing a value for the
parameter defining a critical rate of change $k_c$ beyond which ER cannot occur.
To do so, we use an approximate model with constant environmental change
$\theta(t)=\tilde{k}t$ given $\tilde{k}$ constant in time. Then the
model is identical to a simplified version of B\"{u}rger and Lynch (1995)
presented in Kopp and Matuszewski (2013), and the population reaches a % MLB edit: added comma after Kopp & Matuszewski citation
dynamic equilibrium where the trait lags the optimum by the value
$\tilde k (\sigma_a^2(N) + \omega^2 + \sigma_e^2)/\sigma_a^2(N)$. Using % MLB edit: added missing \ before last \sigma
this, Kopp and Matuszewski (2013) calculate the value of $\tilde{k}$ at
self-replacement such that for any $\tilde k>k_c$ the population is
below replacement (i.e. $\bar \lambda < 1$) and the population will
decline, as
\begin{equation}\label{E:kc}
k_c(N) = \sigma_a^2(N) \sqrt{ \frac{2 \log{\left(R_0 \sqrt{s(N)} \right)}}{\sigma_a^2(N) + \omega^2 + \sigma_e^2}}. % MLB edit: 0 subscripted for R_0
\end{equation}
Here, we still employ the SHC approximation, such that the population size affects the critical
% MLB edit: addition to explain the 500's below; note that this defines a new parameter, added to Table 1:
rate $k_c(N)$, which thus should be computed for the minimum population size reached during ER.
For this, we use a population size, $N=N_{c,g}$ below which negative
factors beyond demographic stochasticity (e.g., mutational meltdown) may cause
rapid population extinction.  % MLB comment: please rephrase as appropriate; I wasn't sure where the 500 came from

\subsubsection{\texorpdfstring{The control: improving population growth
\emph{in
situ}}{The control: improving population growth in situ}}\label{the-control-improving-population-growth-in-situ}

We consider a control that temporarily modifies the population growth
rate \emph{in situ}, resulting in changes in population dynamics and
costs to the manager. If the control increases the population by a
factor $v(t) \ge 1$ at each time $t$ simultaneous with natural production $R_0$ then we replace the population size
$N(t)$ with  $N(t)v(t)$ in eq.~\eqref{E:lambda},
and the population dynamics % MLB edit: changed \textasciitilde{} to ~
with the intervention are
\begin{equation}\label{E:popdyn}
N(t + 1) = N(t) v(t) \bar \lambda(t, v(t) N(t)).
\end{equation}
The mean trait dynamics (eq.~\eqref{E:arecursion2})  % MLB edit: changed \textasciitilde{} to ~
are unchanged.

We assume that interventions incur costs $c(v(t))$ that scale % MLB edit: added $$ around c(v(t))
quadratically with the proportional increase in the growth rate.
We also consider a yearly budget constraint.

\subsection{Statement and analysis of the control problem}\label{statement-and-analysis-of-the-control-problem}

The control problem is to minimize costs $c(v(t))$ while
avoiding population sizes below a critically low level, $N_{c,s}$,
assuming the
growth rate eq.~\eqref{E:lambda}
determines the biological dynamics, values of the control
within the feasible set $v(t) \in \Omega$, and with discount rate $\Delta$
across the time horizon $T$,

\begin{subequations}
\begin{align}\label{E:optpop}
 &{\rm min}_\Omega \; \sum_t c(v(t))\left(\frac{1}{1+\Delta}\right)^t \quad \forall t \le
 T\\
  & \textrm{subject to } N(t) \ge N_{c,s}, \; v(t) \ge 1,
\end{align}
\end{subequations}
where the dynamics of population $N(t)$ are defined in
eq.~\eqref{E:popdyn}.

To analyze the discrete-time optimal control problem
eq.~\eqref{E:optpop} we specify concrete functional forms for the costs
and add constraints based on the population dynamics. For a control $v(t)$
we assume a simple cost function
$c(v(t)) = \log v(t)^2$, which
% some % MLB edit: deleted some
results in a cost of 0 when $v(t) = 1$ and quadratic costs for % MLB edit (x2): changed "in in" to "in a", changed "on for" to "for"
log-scale intervention $u(t) = \log v(t)$. The objective at each time
(neglecting discounting) is then $u(t)^2$.
We also let
$x(t) = \log N(t)$, and denote
the log-scale initial population size as
$x_{init}=\log N(0)$, which enters the problem as an equality constraint at time 0.
We denote the log-scale critical population size as $x_c = \log N_{c,s}$,
which enters the problem as an inequality constraint at each time.
In log scale, the recursion for population growth from eq.~\eqref{E:popdyn}
is $x(t+1) = u(t) + x(t) + \log \bar \lambda(t, \exp(x(t) + u(t)))$, these dynamics
enter the problem as an equality constraint at each time.
Finally, we assume a budget constraint with a constant budget
$b$ within each year, which imposes another inequality constraint at each time.
Accounting for all of this, the constrained control problem is
\begin{subequations}\label{E:control}
\begin{align}
\min_{u(t)} \sum^{T-1}_{t=0} u(t)^2 \left (\frac{1}{1 + \Delta} \right )^t &\quad \textrm{subject to} \\
x(t + 1) - x(t) &= u(t) + \log \bar \lambda \left(t, \exp(u(t) + x(t)) \right) \quad t = 0,1, \dots T\\
x(0) &= x_{init}\\
0 &\le u(t) \quad t = 0,1, \dots T - 1 \\
u(t)^2 &\le b \quad t = 0,1, \dots T - 1 \\
x(t) &\ge x_c \quad t = 0,1, \dots T.
\end{align}
\end{subequations}

\subsubsection{Parameter choices and assumptions}\label{parameters}

To model a situation where the population is initially declining but
could eventually recover (albeit having experience populations too low
to persist), we assume $ t_\text{safe}$, the time at which the rate % MLB edit: changed $ to ( or ) bracketing the equation
of environmental change $k(t)$ in eq.~\eqref{E:kappa} transitions from
being greater than to being equal to the rate at which ER is possible,
$k_c(N_{c,g})$, occurs within the time horizon, i.e., % MLB edit: change 500 to the new parameter N_{c,k}, set to =500 in the table
$t_\text{safe} \le T$.

% MLB edit: addition to have some justification of parameter value choices
We chose the biological parameters to start in a space where, without intervention, the population initially declines to a low population size susceptible to stochasticity but not to deterministic extinction, as that is the parameter space where our central questions on the effects of intervention on ER are relevant.  % MLB comment: please make sure this is correct and edit as necessary
We also assume that the population initially is experiencing a sustainable rate
of environmental change.
See Table \ref{tb:parms} for all default parameter values used. In
addition to analyzing the optimal path of investment in intervention for
these default values, we compare the optimal path under varying
density-dependence $K= 10,000$, $K=15,000$, or $K=20,000$ %,
% MLB edit: additions here and below to remind the reader what questions each of these analyses test in the language of the questions posed in the Introduction
to explore the effect of population regulation, and
a discount rate of $\Delta =$ 0 or 0.025 %,
 and a budget of $b=$ 0.01 or 0.02
to explore the effects of economic factors.
%%
% MLB edits: rearrangements and deletions because I don't think the rate of change affects the initialization, just the initial genetic mean relative to the optimal phenotype
% JDA: assumptions about the intial rate of change do affect the carrying
% because they affect \bar{W}(0)
In all cases,
the initial genetic mean is the initial optimal phenotype \(a_0=\theta(0)=0\) and
the initial population size is set equal to the
equilibrium population size with \(K=10,000\)
accounting for variance load  \(N(0)=\bar{W}(0)(R_0-1)K\)
under the assumption that the environment is already changing at a rate $\kappa_\text{min}$
(this results in $N(0) \approx 3600$).
% MLB comment: is this re-phrasing correct?  I was  confused by K=10,000 and K=15,000 at different places.
% and assuming the
%environment is already changing at a rate $\kappa_\text{min} k_c(N_{c,k})$
%(this results in $N(0) \approx 3600$, which is roughly 66\% of the
% equilibrium population size with $K=15,000$), and the initial genetic
%mean is the initial optimal phenotype $a_0=\theta(0)=0$.
%%

% MLB edit: moved some rows so all parts of the biological model first appear together (and kappa's next to t_safe because together they define the environmental change), then the control problem
\begin{table}[hbt]
\caption{Parameters and default values.}\label{tb:parms}
\begin{tabular}{llp{4in}}
\hline
Parameter & Default Value & Description\\ \hline
$R_0$ & 1.5 & Number of offspring per individual \\
$K$ & 15,000 & Carrying capacity \\
$\omega^2$ & 50 & Selectional variance (inverse of selection strength) \\
$\alpha^2$ & 0.05 & Genetic effect size variance of a new mutation \\
$V_m$ & 0.001 & Mutational variance \\
$\sigma_e^2$ & 0.5 & Environmental contribution to phenotypic variance \\
$t_\text{safe}$ & 20 & Time at which the rate of environmental change slows to a value where ER can occur \\
$\kappa_0$ & 2.5 & Maximum rate of change (multiplied by $k_c(N_{c,g})$)\\ % MLB edit: changed 500 to N_{k,s}
$\kappa_\text{min}$ & 0.95 & Minimum range of change (multiplied by $k_c(N_{c,g})$) \\ % MLB edit (x2): changed 500 to N_{k,s}, added "minimum rate of change"
% MLB edit: added new parameter noted above
$N_{c,g}$ & 500 & Population size used for calculating critical rate of change $k_c$\\
$N_{c,s}$ & 1,000 & Critical population size for extinction risk due to stochastic factors\\
$\Delta$ & 0.025 & Discount rate\\
$b$ & 0.01 & Annual budget \\
\hline
\end{tabular}
\end{table}

\subsubsection{Model analysis}\label{model-analysis}

We numerically analyzed the system \eqref{E:control} with augmented
Lagrangian minimization (Birgin and Mart\'{i}nez 2008) as implemented in the
\texttt{NLOPT} library (Johnson 2016). This requires restating the
problem as a constrained discrete-time optimal control problem (see, e.g., Chow 1997), with
eq.~\eqref{E:control}a as the objective to minimize, eqs.~\eqref{E:control}b and \eqref{E:control}c as equality
constraints and eqs.~\eqref{E:control}d, \eqref{E:control}e, and \eqref{E:control}f as inequality constraints.
See the supplementary methods (Online Resource 1) for code.
For all parameter combinations, we set the initial control and
population to a path found using a zero discount rate
and a large number of iterations. For global optimization algorithms such as the
one we employ, convergence is difficult to assess in general. For a convergence criterion,
we considered a path optimal if the solver consistently converged upon it with an increasing number of
iterations; see Appendix \ref{A:convergence}.

\section{Results}\label{results}

Given our
% MLB edit: addition
choice of parameter space and
assumption that the rate of environmental change starts
greater than, and eventually shows to, a value where ER is possible
(Figure \ref{figureenv}a), without intervention population growth is
initially negative and population size falls below a critically
population vulnerable to extinction (Figure \ref{figureenv}b).
Eventually, as environmental change slows, population population growth
will become positive (Figure \ref{figureenv}b), with a U-shaped
demographic trajectory analogous to ER models with sudden environmental
shifts (Gomulkiewicz and Holt 1995; Carlson et al. 2014).

\subsection{Optimal investment trajectory and
indicators}\label{optimal-investment-trajectory-and-indicators}

\begin{figure}[p!]
  \centering
  \includegraphics[width=\textwidth]{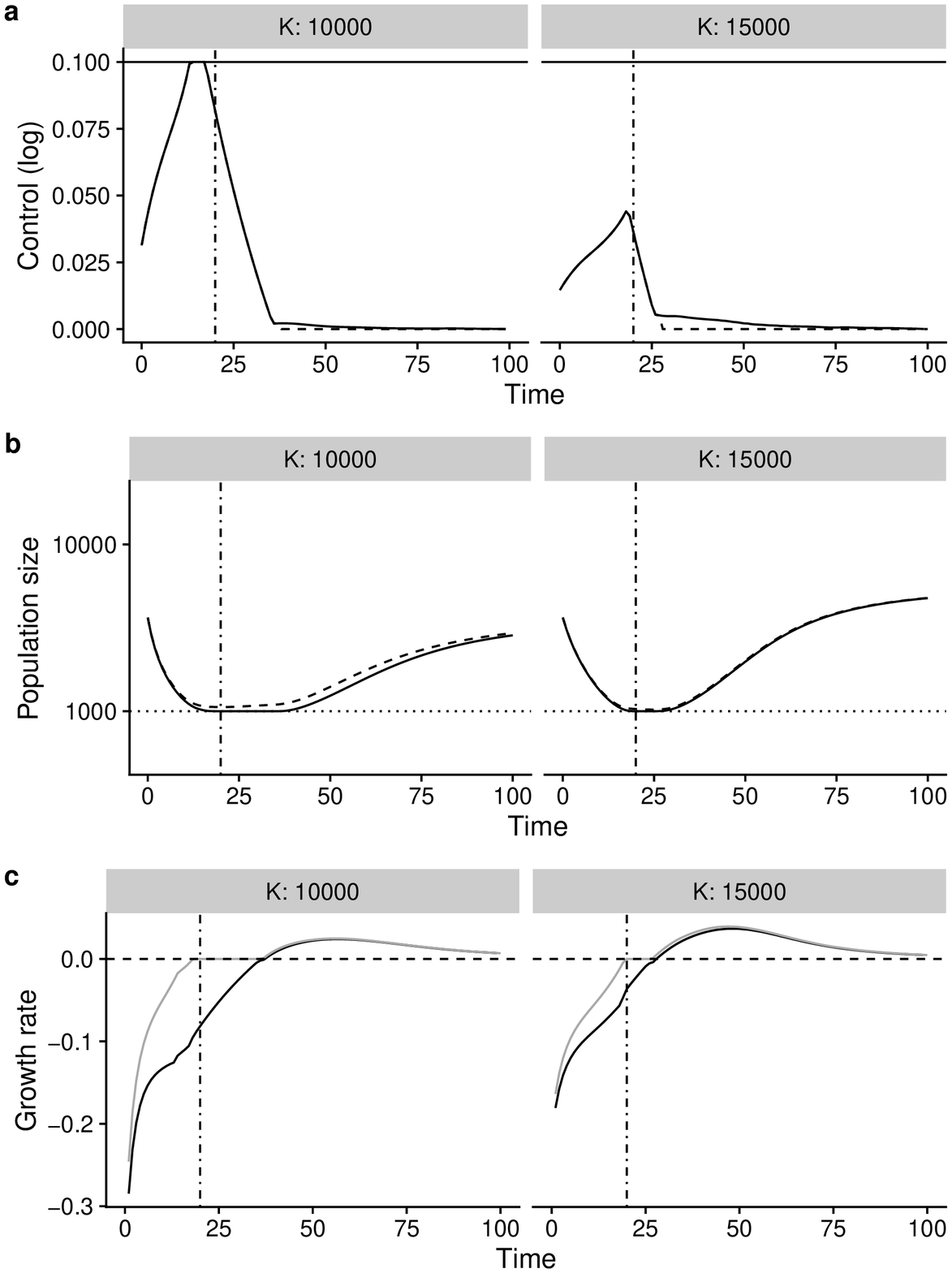}
	\caption[Control paths, population trajectories, and growth rate for varying
carrying capacity and economic factors.]{Details of the (a) optimal investment path, (b) population size, and (c)
population growth rate ($\log\bar{\lambda}$) under the default
parameter values (Table \ref{tb:parms}) with varying $K$. The strategy that
minimizes costs to intervention while avoiding a population size below
a threshold vulnerable to extinction (horizontal dotted line in B)
results in an initial increase in investment, which peaks then decreases
(a) in the same year that the growth rate including the intervention
(gray line) transitions from negative to zero (stable; C); note that
without the intervention population growth rate would still be negative
(black line in C). Investment ramps down after the time
$t_\text{safe}$ (vertical dash-dot line) when the rate of
environmental change equals the critical rate where evolutionary rescue
can occur, i.e.~population growth can become positive without
intervention (Figure \ref{figureenv}).}
	\label{figureout}
\end{figure}

The optimal trajectory for investment in intervention initially
increases quickly, with investment peaking at or before the time when
the population size reaches the minimum acceptable population size
(Figure \ref{figureout}). Notably, investment increases even as the
population is declining under the management intervention (Figure
\ref{figureout}c). Thus, in this case, declining population under a
management intervention does not imply that the strategy is non-optimal. % MLB edit: added "that" after imply

Optimal investment then begins to slow in the year that the population
growth rate, including effects of intervention (gray line in Figure
\ref{figureout}c), transitions from population decline to stable.
Investment reaches a very low level once population growth rate would be
positive without the effects of intervention (Figure \ref{figureout}c); % MLB edit: added ;
this occurs with some delay after the rate of change decreases to the
critical rate $k_c$. Once that occurs, at time $t_\text{safe}$, the
rate of environmental change is still positive, but at a rate slow
enough for evolutionary rescue to occur if it were constant; however,
intervention is still needed after $t_\text{safe}$ to reduce lag
between the population mean trait and the optimal trait to a sustainable
magnitude.

\subsection{Factors that influence the optimal trajectory of
investment}\label{factors-that-influence-the-optimal-trajectory-of-investment}

\begin{figure}[p!]
  \centering
  \includegraphics[width=\textwidth]{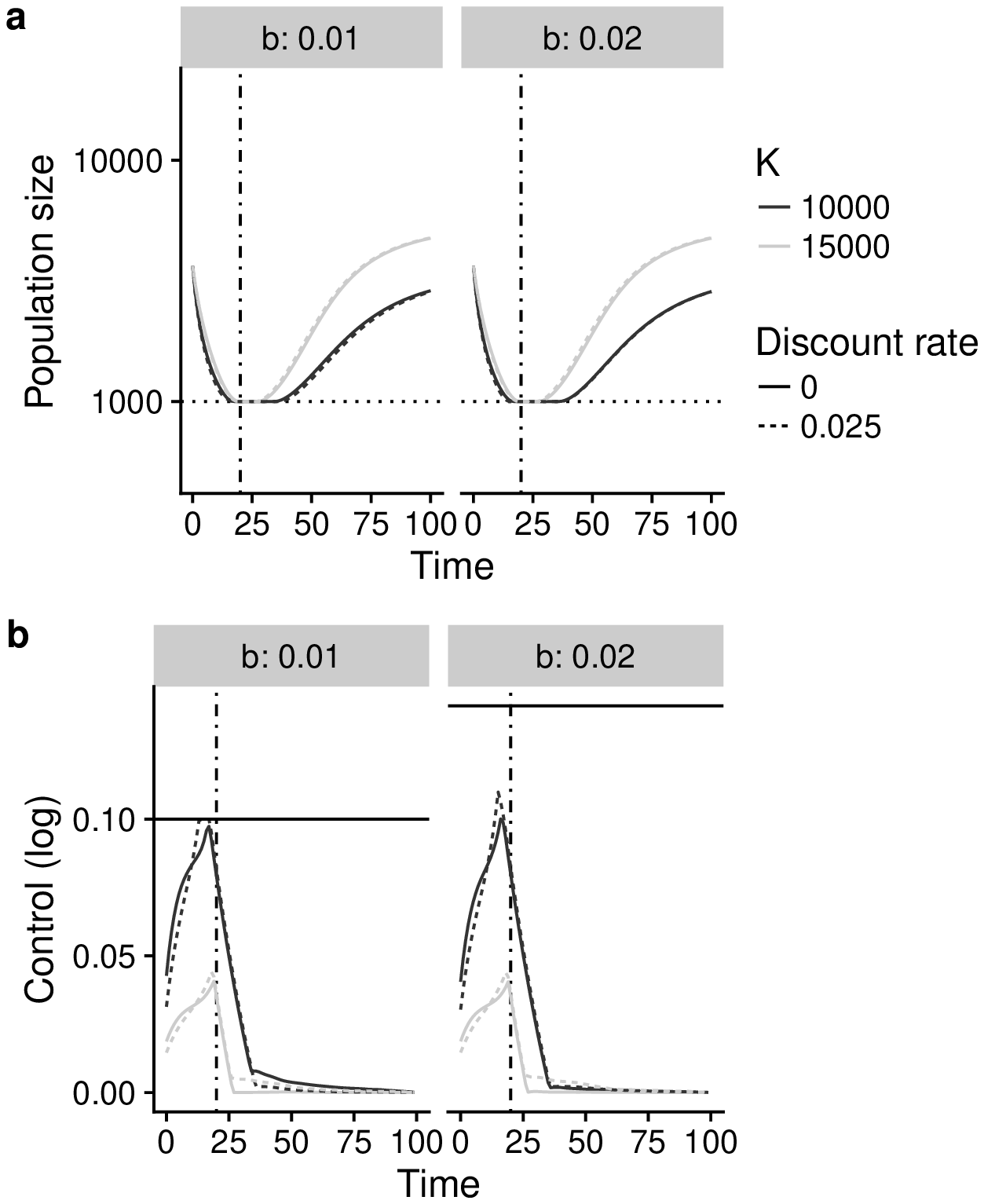}
	\caption[Optimal control paths and effect on population trajectories with varying
carrying capacity.]{Optimal control with varying carrying capacity (varying greyscale) and
discount rate (varying linetype) for two yearly budgets. (a) Population
sizes under the optimal path of investment in intervention, which
includes the constraint of not allowing the population to fall below a
size considered vulnerable to extinction (horizontal dotted line). (b)
Optimal investment trajectory %,
% MLB edit: addition to make sure we describe all lines
relative to the budget constraint (horizontal line),
where investment ends when % MLB edit: addition to clarify
population growth is positive, lagged after
the rate of
change decreasing to the critical rate where evolutionary rescue can
occur at $t_\text{safe}$ (vertical dash-dot line). Discount rates have
little effect relative to that of carrying capacity. A carrying capacity
closer to the initial population size (which corresponds to
$K=10,000$) leads to initially-steeper population declines and earlier
peak investment. The larger budget ($b=0.02$ per year) never
constrains the solution. All solutions assume decelerating environmental
change as in Figure \ref{figureenv}a where the rate of environmental
change decreases to the critical rate where evolutionary rescue can
occur at $t_\text{safe}$ (vertical dash-dot line).}
	\label{figurevaryKdeltab}
\end{figure}

A carrying capacity further from the initial population size favors lower % MLB edits: framed in terms of direction of management action (increasing carrying capacity, previously phrased in terms of what happens with lower carrying capacity)
investment overall and shifts investment later in time (Figure
\ref{figurevaryKdeltab}). %; initial population size corresponds to lightest gray line).
Compared to carrying capacity, the economic factors of
discounting and budget constraints had weaker effects on the amount of
investment in intervention (Figure \ref{figurevaryKdeltab}-\ref{figurevaryKdelta0b}). Greater
discounting favors investing later in time (Figure
\ref{figurevaryKdelta0b}) and weakens the need to ramp investment down to
zero after positive population growth is achieved (with zero discounting
investment goes to practically zero at this point; see Appendix, Figure
\ref{figurevaryKdelta0b}).
%%
% MLB comment: do or cut
% Extending the
% timescale of optimization, $T$, did not change the pattern of
% investment for fixed $t_\text{safe}$ (TODO).
% * <mlbaskett@ucdavis.edu> 2018-08-07T18:08:38.216Z:
%
% This text showed up in the "diff" version, but looks like it isn't in the version without differences highlighted -- make sure that it doesn't show up in the final version.
% > % Extending the
% > % timescale of optimization, $T$, did not change the pattern of
% > % investment for fixed $t_\text{safe}$ (TODO).
%
% ^.
%%
% A constraining budget
% favors a steeper increase in spending early, until the constraint is met
% (Figure \ref{figurevaryKdeltab}). This results in the counter-intuitive
% pattern where with a constraining budget the population actually takes
% longer to reach the minimum size than an non-constraining budget (Figure
% \ref{figurevaryKdeltab}b).
%JDA: note this effect is very weak and we shouldn't discuss it but teh reason
%it's counter-intuitive is that one might think greater budgets would equal
%slower declines in the managed resource

\section{Discussion}\label{discussion}

We find that, with decelerating change, short-term investment in
enhancing population growth can reduce extinction risk to allow for a
combination of evolution and global-scale mitigation (resulting in
deceleration of the optimum) to lead to long-term persistence. This
occurs because at the time investment is stopped, the rate of change is
within the population's tolerance limits (see, e.g., B\"{u}rger and Lynch
1995). Optimal investment trajectories to conserving populations in the
face of global stressors may initially mean doubling down on what
appears to be a failing strategy due to ongoing population decline
(Figure \ref{figureout}a,b). Mumby et al. (2017) provide a similar
example where a declining system state is not a signal of improper
management. In their analysis of coral reef management under climate
change, they point out if managers and the public consider the unmanaged
(or non-optimally managed) counterfactual scenario then this can alter
perceptions of management utility. Such analyses are necessary to evaluate the
effectivness of management and distinguish between those strategies that are actually
failing and those which are optimial but still result in declines; our results demonstrate that
such exercises may be needed to avoid a crisis of motivation when
managing populations that are capable of evolutionary rescue.

In contrast to the trend or status of population size, under optimal
management the trend in population growth rate (including management's
effect on population growth) reliably increases, at first becoming less
negative and eventually leveling out to stable then increasing to
persistence (positive population growth; Figure \ref{figureout}c). This
indicates that trend in growth rate may provide a reliable signal of
management efficacy as compared to the trend in population size. These
same observations imply that timing of assessment matters: assessing the
effect of an intervention prematurely may lead managers to dismiss what
would be a successful strategy in the long run.

Overall, the optimal investment trajectory of initially increasing,
then, as population growth stabilizes, decreasing, to stop when
% MLB edit: updated carryover from previous version, before we found the error in the derivation
% environmental change is slow enough for evolutionary rescue to occur,
population growth is positive,
is surprisingly robust to a wide array of economic assumptions and
parameters, both qualitatively and quantitatively (Figure
\ref{figurevaryKdeltab}). Note that this unimodal investment trajectory
is analogous to that in Lampert and Hastings (2014), focused on the
optimal investment schedule for restoration to accelerate the recovery
of a degraded system in a stable environment (without evolution). Much
like the cessation of investment when evolutionary rescue can occur
naturally in our model, the optimal investment trajectory in Lampert and
Hastings (2014) ceases after at an ``economic restoration threshold'',
before full recovery has occurred.
Both Lampert and Hastings (2014) and our study are examples of a phenomenon
that is likely more general in conservation decisionmaking: optimal managment
involves investing to a point when natural processes can complete recovery.

Carrying capacities closer to the initial population size led to earlier
and greater investment in population enhancement (Figure
\ref{figurevaryKdeltab}), which indicates a significant role of
density-dependent suppression of population growth even for declining
populations. This result reflects the fact that per-capita reproduction decreases
as the population approaches the carrying capacity, and again points to population growth serving as a
more useful indicator than population size: while a population size near
carrying capacity might, based on intuition, be considered to be not yet
in need of support, the faster initial decline (due to stronger density
dependence in combination with rapid environmental change) means that it
actually requires greater initial intervention. In addition, this result
indicates that a separate investment to permanently enhance carrying
capacity, such as through restoration, can significantly reduce the
investment necessary in short-term population enhancement, such as
through captive rearing or breeding. A key next step in this analysis
would be to analyze the optimal investment across actions with long-term
and short-term effects; note that, unless the action with long-term
effects enhances population growth rather than carrying capacity,
investment in short-term population enhancement will always be necessary
under our model assumptions given rapid environmental change leading to
initial population declines.

\subsection{Applications}

Our model provides a generic representation of cases of systems where climate change
might threaten near-term persistence and interventions to increase population size during such a period are feasible.
Examples include climate change-driven changes to rivers treatening Pacific salmon
Pacific salmon (Crozier et al 2008) whose populations can be supplemented via hatcheries (Naish et al 2008),
and climate threats to the persistence of tropical corals (Bellwood et al 2004) whose populations can be supplemented via
"coral gardening" (Lirman and Schopmeyer 2016).

Direct application of our model to one of these cases would require empirical knowledge of
both genetic potential and the change in the environmental optimum, as well as other biological parameters.
Although such estimates of genetic parameters are often available (see, e.g., Carlson et al 2014)
estimates of environmental optima are rare, but critical for predicting evolutionary responses
to environmental change (Chevin et al 2010; Chevin et al 2017).
Our analysis demonstrates how such predictions might be used by managers; the
next step is to develop parameter estimates and models for specific settings.
Such case-specific models would need to address not only biological parameters
but policy choices, for example the quasi-extinction threshold, $N_{c,s}$.
In fact, even the use of a threshold is a choice.
For some cases, an alternative model where a explicit value is placed on existence
of the population my be preferable.

\subsection{Assumptions and analytical
choices}\label{assumptions-and-analytical-choices}

As with any model, our model makes a number of simplifying assumptions
for tractability. For example, we use a generic form of population
enhancement that temporarily increases growth rate, which we associate
with actions such as resource supplementation, head-starting, or captive
breeding. As noted in the Introduction, such actions might incur
unintended consequences such as domestication and reduced fitness, which
we ignore with our assumption that the genetic dynamics (dynamics of
$\bar{a}_t$) are independent of the intervention. For example,
reductions in wild fitness occurs rapidly in Pacific salmon reared in
hatcheries (Araki et al. 2008; reductions can occur within one
generation, Christie et al. 2012). Incorporating such unintended fitness
consequences of captive rearing would likely delay the evolutionary
response in our model and therefore might increase the duration of
intervention necessary given our constraint of maintaining a population
size above a critical threshold, depending on how much an increase in
program duration intensifies domestication selection. Quantitative
genetic models indicate that one potential approach to reducing such
unintended fitness consequences is to consistently target a combination
of captive-reared and wild-reared individuals in the captive environment
(Ford 2002; as opposed to captive-reared only, Baskett and Waples 2013).
Alternatively, careful management of breeding's effects on genetic
variance in trait and fitness might prove an accelerator for evolution
and be purposely used (as in ``adaptive provenancing'' \emph{sensu}
Weeks et al. 2011; or ``assisted evolution'' \emph{sensu} Oppen et al.
2015), where the balance between domestication and assisted evolution
effects would determine the efficacy of this approach.

One major omission from our modeled scenario is phenotypic plasticity.
When phenotypes plastically respond to environmental change, this can
facilitate adaptation to a changing environment (Chevin et al. 2010) and
thus evolutionary rescue; the relationship between the environmental cue
that affects phenotype and the environment of selection, however, is
critical for determining whether plasticity increases the chances of
evolutionary rescue (Ashander et al. 2016). However accounting for
plasticity may be important in understanding the effects of climate
change, as much of the response in traits observed to date owes to
plasticity (Meril\"{a} and Hendry 2014). This may be especially true for
species with complex life cycles involving many transitions between
environments (e.g., Pacific salmon, Crozier et al. 2008).

Our modeled intervention to increase short-term population growth
assumes immediate effect. In reality, many interventions, such as
habitat restoration or removal of stressors like invasive species, might
have delayed effects and require intervention over multiple years for a
permanent effect to occur (Myers et al. 2000; Borja et al. 2010). In a
discrete-time formulation such as ours, delays like this would likely
result in greater investment earlier in time. These and other subtleties
warrant investigation in future work on the interaction between
microevolution and restoration, a topic of increasing import given
climate change (Rice and Emery 2003).

In our analysis, we rely on a threshold population size $N_{c,s}$ to
indicate extinction risk to factors such as demographic stochasticity,
environmental stochasticity, Allee effects, inbreeding, and genetic
drift. Although this approach is common, and may seem conservative
(Gomulkiewicz and Holt 1995), it may mislead. Explicit analyses of
demographic and genetic stochasticity can more effectively describe how
extinction risk varies with factors such as genetic variance and
indicate that minimum population size might better predict extinction
risk than time below a threshold (Boulding and Hay 2001). However, for
applications it is more common to set management goals in terms of
population size as compared to actual extinction risk (Flather et al.
2011). In part, this may be because population size is easier to
quantify than risk.

For our population dynamics, we further assume a saturating,
Beverton-Holt (1957) form of density-dependent regulation, which ignores
the potential for overcompensation (i.e., a decline, rather than
saturation, at large population sizes, as in Ricker density dependence).
Strong overcompensation would likely delay the optimal initial
investment until after some population decline has occurred, such that
enhanced population growth would not increase the population size beyond
the overcompensatory level where large-scale declines would then occur.
Analogously, an initial action to permanently increase carrying capacity
and therefore weaken density dependence might have even a stronger
effect under overcompensatory density dependence.
However, we examined only a single
life-cycle ordering (reproduction, density-dependence, viability
selection), which corresponds to hard selection (Wallace 1975).
Viability selection occurring before, rather than after, density
dependence would likely reduce the role of increasing carrying capacity
in decreasing the amount of investment necessary.
Further, we examined only non-overlapping generations without age structure.
The response of such populations is an intersting topic for future work,
as it is unclear whether they would respond more or less rapidly than the
case of non-overlapping generations studied here.
On the one hand, overlapping generations with age structure can increase the maintenance of genetic variation (Ellner and Hairston 1994),
and greater genetic variation can mean greater adaptive capacity and therefore more rapid evolutionary response.
On the other hand, generation time is longer in such populations, resulting in slower evolutionary response.

We relied on standard assumptions for quantitative genetic models, which
include a large number of loci contributing additively to a trait with a
normal distribution % and constant genetic variance % MLB edit: deleted because genetic variance isn't constant wit hthe hous-of-cards approximation
(Lande 1976; Lande 1982). Such
assumptions typically have minor effects on the predicted evolutionary
trajectory (Turelli and Barton 1994).
% MLB edit: changed to reduce confusion about what we did vs. didn't model
%, except for the assumption of
%constant variance which is unlikely to hold for a population declining
%to small size, where effects of demographic stochasticity and drift may
%strongly reduce variance. To account for this,
We did account for the effect of population size on genetic variance, where
we used the stochastic
house-of-cards (SHC) approximation as in B\"{u}rger and Lynch (1995).  % which
% predicts the genetic variance at a mutation-selection-drift balance. % MLB edit: moved this up when we first mention the SCH approximation
This captures the effect of how small population sizes will lower
genetic variance, thus reducing the capacity for evolutionary rescue
(Lynch and Lande 1993; B\"{u}rger and Lynch 1995; Gomulkiewicz and Holt
1995; Carlson et al. 2014) and therefore likely increase the amount and
duration of investment necessary. Although, as Kopp and Matuszewski
(2013) point out, the SHC approximation does not account for the effect
of directional selection on increasing genetic variance, both this % MLB comment: shouldn't "increasing" be "decreasing" here?
% No see discussion on pp 174 of Kopp which reads in part
% [the SHC], is still not the whole story,
% because once the optimum starts moving, r 2 g is expected to
% increase. This increase is mainly due to the rise in fre-
% quency of previously rare alleles, and it is strongest in large
% populations (B€
% urger 1999): for example, under standard
% values of mutational and selectional parameters, r 2 g
% increases up to 4-fold in populations with N e > 5000. In
% contrast, selection has little impact on r 2 g if N e < 200300
% (B€
% urger 1999), which might explain why genetic variances
% usually do not increase in artificial selection experiments,
% as noted by (Johnson and Barton 2005). A useful upper
% limit for the genetic variance in small populations
% urger and Lynch 1995) is the neutral expecta-
% (N e < 500, B€
% tion 2V m N e , where V m is the input of genetic variance from
% new mutations (a typical value is V m 1⁄4 0:001r 2 e , Lande
% 1976a; Lynch 1988). In summary, evolution of the genetic
% variance may increase the prospects of population survival,
% but mostly in large populations. It should be noted,
% though, that the increase in variance takes time and may
% come too late for populations subject to strong environ-
% mental change.
effect and the mutation-selection-drift balance modeled by the SHC occur
only after some transient period; the SHC approximation likely captures
the correct overall average effect of declining population size:
reducing genetic variance.

Our major economic assumption is that cost of the intervention is
quadratic in the amount of intervention. In reality, there might be
decreasing costs, i.e., returns to scale, for supplementation programs.
However, for planning initial investments in a program for a small and
declining population, the context we focus on here, such returns may
never be achieved.

\subsection{Evolutionary rescue modeling
frameworks}\label{evolutionary-rescue-modeling-frameworks}

As noted in the introduction, our model of a decelerating optimum is an
intermediate between the typical evolutionary rescue models of either a
sudden environmental shift (Gomulkiewicz and Holt 1995; Carlson et al.
2014) or an ongoing moving optimum (Lynch and Lande 1993; B\"{u}rger and
Lynch 1995). Because we assume that the rate of environmental change is
initially greater than the critical rate of change for evolutionary
rescue to occur, without (and even with) intervention, we find a
U-shaped population trajectory of initially decreasing, then increasing,
population size (Figure \ref{figureenv}b), commonly associated with
models that have sudden environmental shifts. As compared to other
moving-optimum models, which typically use criteria for rescue that are
conservative and imply that, when evolutionary rescue occurs, population
size never declines (Kopp and Matuszewski 2013), we present a more
realistic representation of environmental change such as climate change
(albeit one that does not yet include effects on climate variability),
while still constructing a generic model as compared to system-specific
models of evolutionary response to local-scale climate trajectories
(e.g., Baskett et al. 2009, 2010; Sinervo et al. 2010; Reed et al.
2011). Therefore, the decelerating optimum model illustrates a general approach to exploring
the interaction between mitigation (management to reduce the rate of
change) and adaptation (management to enhance the capacity for local
systems to respond to change) in promoting evolutionary rescue and
population persistence under climate change.

\begin{acknowledgements}
Thanks to Alan Hastings and Michael Turelli for feedback on an earlier
draft. Special thanks to Lisa Crozier, with whom a conversation planted
the seed of the question investigated here. Also special thanks to Alan
Hastings, for whom this special issue is in honor, for his generosity as
both a mentor and collaborator, as well as the continued inspiration
from his research program to engage in multidisciplinary research and
apply models to conservation challenges. This project was funded by the
REACH IGERT as a Bridge RA (NSF DGE-0801430 to P.I. Strauss) to JA.
\end{acknowledgements}

\appendix

\section{Initialization and convergence of optimal paths}\label{A:convergence}
\begin{figure}[p!]
  \centering
  \includegraphics[width=\textwidth]{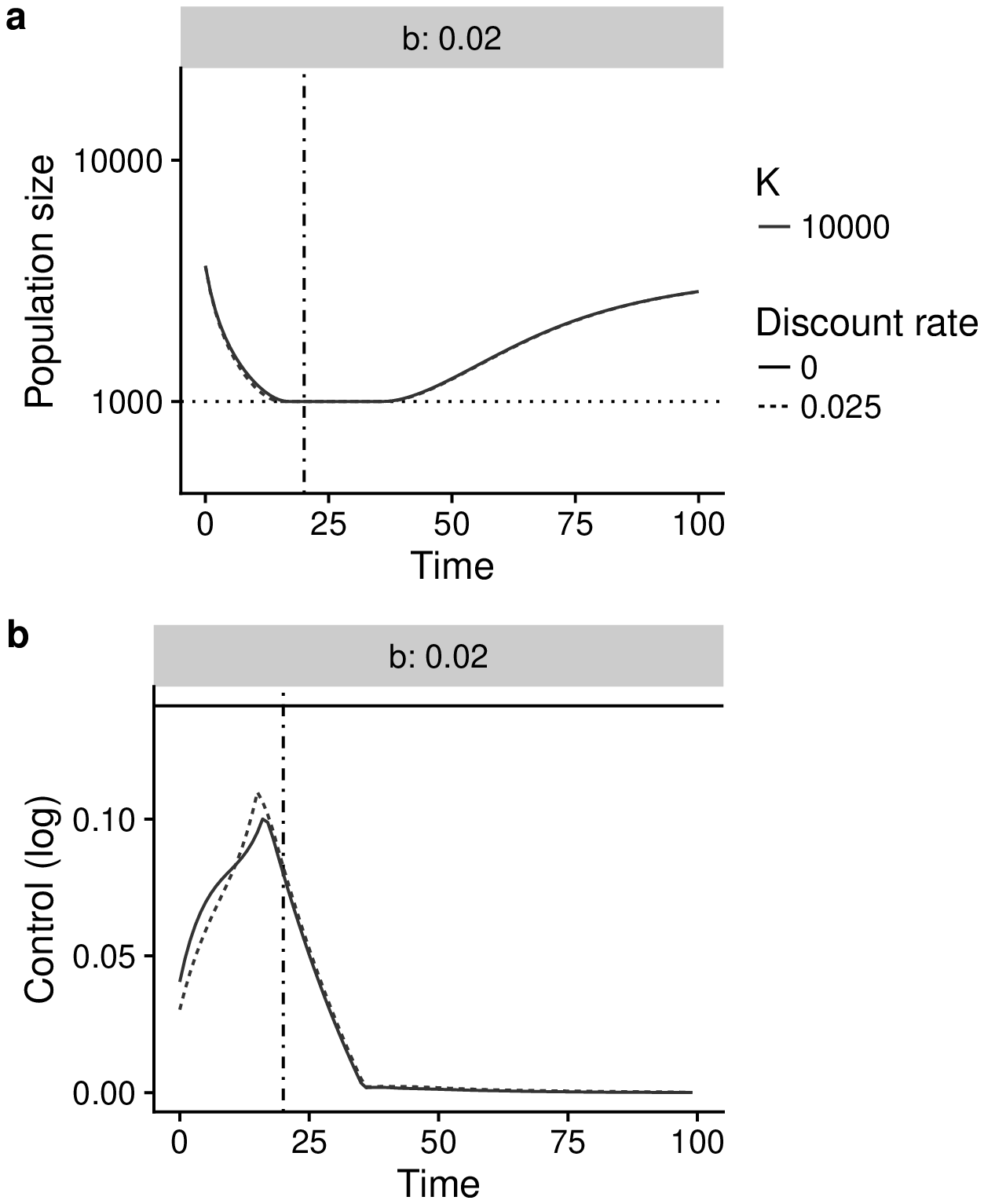}
  \caption[Optimal control paths and effect on population trajectories with zero discount.]{
	  Optimal control with varying
	  discount rate (varying linetype) including zero discount.
	  The zero-discount path was used to initialize the runs with positive
	  discount.
	  (a) Population
	  sizes under the optimal path of investment in intervention, which
	  includes the constraint of not allowing the population to fall below a
	  size considered vulnerable to extinction (horizontal dotted line). (b)
	  Optimal investment trajectory %,
	  relative to the budget constraint (horizontal line).
	  All solutions assume decelerating environmental
	  change as in Figure \ref{figureenv}a where the rate of environmental
	  change decreases to the critical rate where evolutionary rescue can
	  occur at \(t_\text{safe}\) (vertical dash-dot line).}
	\label{figurevaryKdelta0b}
\end{figure}

\begin{figure}[p!]
  \centering
  \includegraphics[width=\textwidth]{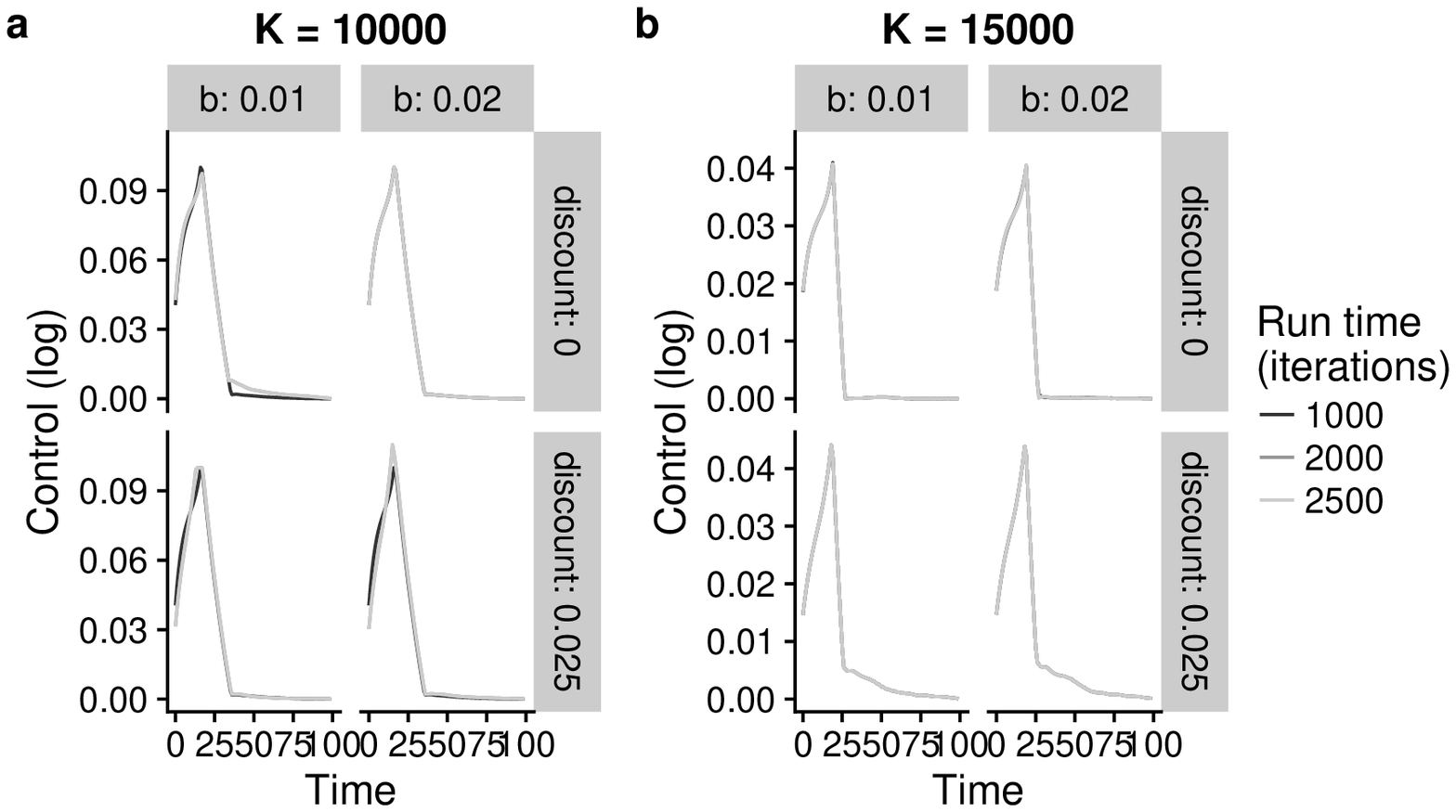}
  \caption[Convergence of optimal control paths]{
	  Optimal paths for increasing run times (lighter greys) to show
	  convergence of control paths for
	  different budgets (columns) and discount rates (subpanel rows) at three
	  carrying capacities corresponding to (a) $K=10,000$, (b) $K=15,000$.
	  There are three runs shown in each panel: 1000, 2000,
	  and 2500 iterations. In most cases the two longest runs resulted in the same
	  path.}\label{figurevaryKdeltavaryruntimecontrol}
\end{figure}

\begin{figure}[p!]
  \centering
  \includegraphics[width=\textwidth]{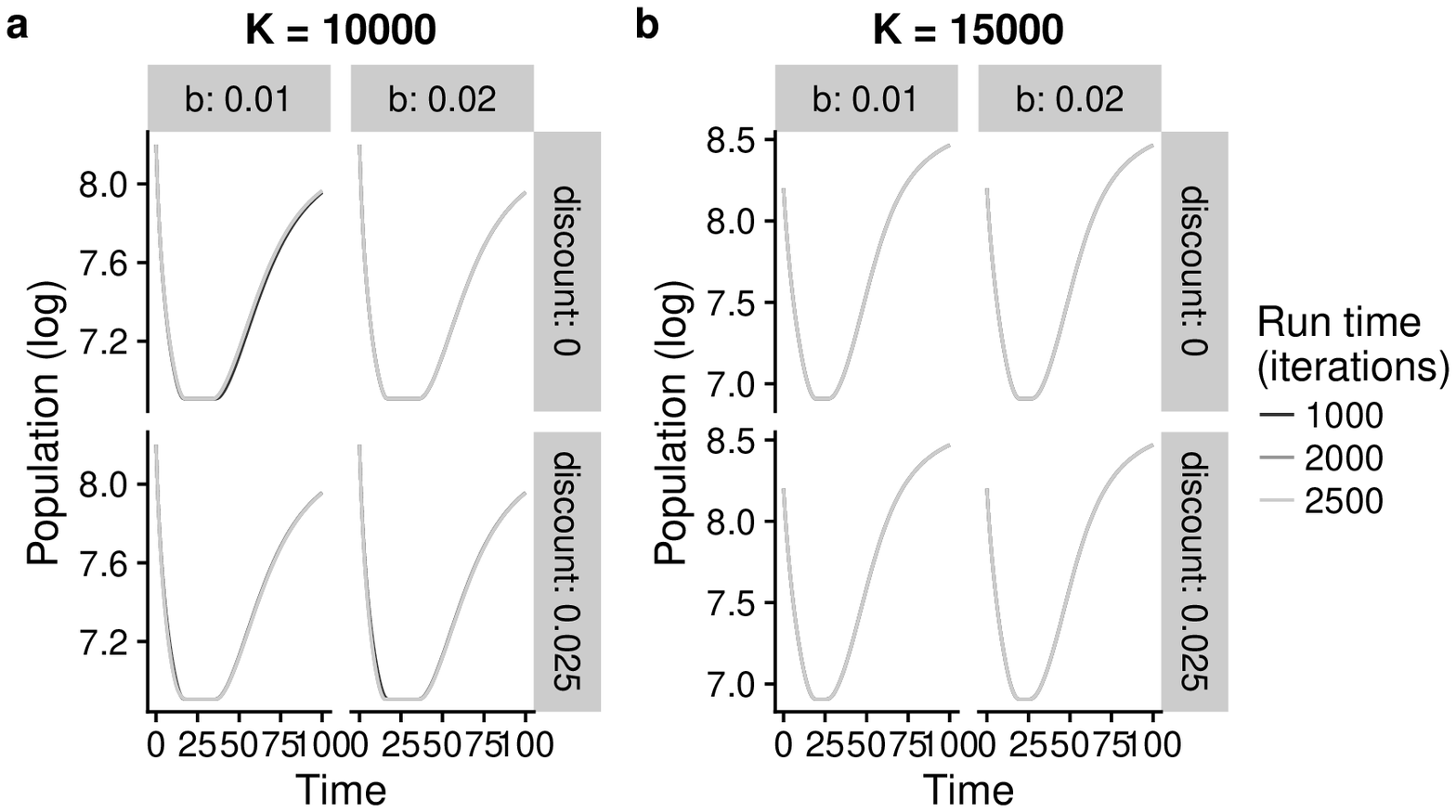}
  \caption[Convergence of effect on population trajectories]{
	  Optimal paths for increasing run times (lighter greys) to show
	  convergence of population trajectories for
	  different budgets (columns) and discount rates (subpanel rows) at three
	  carrying capacities corresponding to (a) $K=10,000$, (b) $K=15,000$.
	  There are three runs shown in each panel: 1000, 2000,
	  and 2500 iterations. In most cases the two longest runs resulted in the same
	  path.}\label{figurevaryKdeltavaryruntimepop}
\end{figure}

We initialized all optimization runs from the
optimal path found with zero discount rate with uniform random initial conditions, and run for 2,500 iterations
(Figure \ref{figurevaryKdelta0b}).
To assess convergence, we re-ran each parameter combination for 1000,
2000, and 2500 iterations.
The longer runs showed consistent paths (Figures
\ref{figurevaryKdeltavaryruntimecontrol},
\ref{figurevaryKdeltavaryruntimepop}),
which is a criterion for convergence recommended for global optimization
algorithms like the augmented Lagrangian method we employed (Johnson 2016).

\section{Code and graphics}
We performed all numerical analyses in \texttt{R} using the \texttt{nloptr}
package to perform optimization and \texttt{dplyr} to manage numeric outputs; we provide R code and metadata for optimal paths in Online Resource 1;
the optimal path used for initial conditions is provided in Online
Resource 2 and the optimal paths for all parameter combinations are provided in
Online Resource 3.
We produced all graphics using R packages \texttt{ggplot2} and \texttt{cowplot}.

\clearpage
\clearpage

\end{document}